\documentclass[twocolumn]{aastex631}

\usepackage{comment}
\usepackage{amssymb}
\usepackage{amsmath}
\usepackage{xspace}
\usepackage{enumitem}
\usepackage{graphicx}
\usepackage{rotating}
\usepackage{color}

\newcommand{\hi}         {\ion{H}{1}\xspace}
\newcommand{\heii}       {\ion{He}{2}\xspace}

\newcommand{\NHI}        {\relax\ifmmode{N_{\rm HI}\xspace} \else {$N_{\rm HI}$}\expandafter\xspace\fi}
\newcommand{\NHIcl}      {\ifmmode{N_{\rm HI,cl}\xspace}\else{$N_{\rm HI,cl}\,$\xspace}\fi}

\newcommand{\kms}        {\ifmmode{\rm \,km\,s^{-1}}\else\,km\,s$^{-1}$\xspace\fi}
\newcommand{\unitNHI}    {\ifmmode{\rm \,cm^{-2}}\else\,cm$^{-2}$\xspace\fi}  
\newcommand{\vexp}       {\relax\ifmmode {v_{\rm exp}} \else {$v_{\rm exp}$}\expandafter\xspace\fi}
\newcommand{\vran}       {\relax\ifmmode {v_{\rm ran}} \else {$v_{\rm ran}$}\expandafter\xspace\fi}
\newcommand{\openangle}       {\relax\ifmmode {\theta_{\rm o}} \else {$\theta_{\rm o}$}\expandafter\xspace\fi}

\definecolor{myblue}{rgb}{0.13, 0.13, 0.55}

\begin{document}


\title{Distribution and Kinematics of \hi through Raman \heii Spectroscopy of NGC~6302}

\correspondingauthor{Hee-Won Lee}
\email{hwlee@sejong.ac.kr}

\author[0000-0002-0112-5900]{Seok-Jun Chang}
\affiliation{Max-Planck-Institut f\"{u}r Astrophysik, Karl-Schwarzschild-Stra$\beta$e 1, 85748 Garching b. M\"{u}nchen, Germany}

\author[0000-0002-1951-7953]{Hee-Won Lee}
\affiliation{Sejong University, 209 Neungdong-ro, Gwangjin-gu, Seoul 05006, Korea}

\author{Jiyu Kim}
\affiliation{Sejong University, 209 Neungdong-ro, Gwangjin-gu, Seoul 05006, Korea}

\author{Yeon-Ho Choi}
\affiliation{Korea Astronomy and Space Science Institute, 776 Daedeokdae-ro, Yuseong-gu, Daejeon 34055, Korea}
\affiliation{University of Science and Technology, 217, Gajeong-ro, Yuseong-gu, Daejeon 34113, Korea}



\begin{abstract}

The young planetary nebula NGC~6302 is known to exhibit Raman-scattered \heii features 
at 6545~\AA\ and 4851~\AA. These features are formed through inelastic scattering of He~II$\lambda\lambda$ 1025
and 972 with hydrogen atoms in the ground state, for which the cross sections are $1.2 \times 10^{-21}$ and $1.4\times 10^{-22} {\rm\ cm^2}$, respectively.
We investigate the spectrum of NGC~6302 archived in the ESO Science Portal. 
Our Gaussian line fitting analysis shows that 
the Raman-scattered \heii features are broader and more redshifted than the hypothetical model
Raman features that would be formed in a cold static \hi medium.
We adopt a simple scattering geometry consisting of a compact \heii emission region
surrounded by a \hi medium to perform Monte Carlo simulations using the radiative transfer code {\it STaRS}.
Our simulations show that the \hi region is characterized by the \hi column density $N_{\rm HI}=3\times 10^{21}{\rm\ cm^{-2}}$ with the random speed component $v_{\rm ran}=10{\rm\ km\ s^{-1}}$ expanding
with a speed $v_{\rm exp}= 13{\rm\ km\ s^{-1}}$ from the \heii emission region. 
Based on our best fit parameters, we estimate the \hi mass of the neutral medium 
$M_{\rm HI} \simeq 1.0\times 10^{-2}\ {\rm M_\odot}$, pointing out the usefulness
of Raman \heii spectroscopy as a tool to trace \hi components.

\end{abstract}

\keywords{Radiative transfer --- Planetary nebulae --- Scattering --- Individual NGC~6302}


\section{Introduction} \label{sec:intro}

NGC~6302 is a young planetary nebula exhibiting a well-known butterfly morphology.
The two main lobes are divided by an equatorial torus composed of atomic, molecular, and dusty material \citep{matsuura05,kastner22}. With high helium and nitrogen abundances, NGC~6302 is classified
as a Type I planetary nebula according to the classification scheme proposed by \cite{peimbert78}.
It belongs to the highest excitation class with prominent
emission lines including N V$\lambda\lambda$1238, 1243, C~IV$\lambda\lambda$1548, 1551
and Ne~VI at $7.7{\rm\ \mu m}$ \citep{feibelman01,pottasch85}.
Many researchers
investigated NGC~6302 for the internal kinematics and various components, including the ionized, atomic, molecular
and dust components. 

\cite{meaburn08} investigated the proper motions of the outflowing knots to propose that the  distance to NGC~6302 is 1.17 kpc. They also derived a kinematic age of 2200 years from their
analysis of the Hubble-type expansion \citep[e.g.,][]{szyszka11}. \cite{dinh08} proposed that the molecular
torus is expanding with a speed of $\sim 15{\rm\ km\ s^{-1}}$ from their measurement using the 
Submillimeter Array. \cite{santander_garcia17} conducted a kinematical
analysis using ALMA data to confirm the Hubble type expansion.

With recent history of significant mass loss, a young planetary nebula is expected to harbor abundant \hi behind the ionization front, which is expanding with respect
to the central hot source. The hyperfine structure 21 cm line is regarded as currently the most
effective spectroscopic tracer of atomic hydrogen. 
The first successful detection of a neutral component in planetary nebulae was made by
\cite{rodriguez82}, who conducted radio observations of NGC~6302 using the Very Large Array. 
CO and \hi components were also detected from radio observations for a number of planetary nebulae \citep[e.g.,][]{gussie95}. 
However, severe confusion
from the Galactic emission prevents one from investigating the distribution and the
kinematics of the \hi components in planetary nebulae. 

For young planetary nebulae, a very unique and useful spectroscopic probe is provided
by the Raman scattering process of far UV line radiation with atomic hydrogen. 
In young planetary nebulae, far UV \heii lines can be
Raman scattered by atomic hydrogen to form broad features blueward of hydrogen Balmer lines
\citep{nussbaumer89}.
The first report of Raman-scattered \heii features was made for the young planetary nebula
NGC~7027 by \cite{pequignot97}, who identified the broad feature at 4852 \AA\ as Raman-scattered \heii. \cite{groves02} found the same feature in NGC~6302 while they investigated
the extinction in the nebula.
Subsequently, Raman-scattered \heii at 6545 \AA\ was detected in the young planetary nebulae IC~5117, NGC~6790, NGC~6881, and NGC~6886 \citep{lee06, kang09,choi20b}.

Raman-scattered \heii features are clearly detected in the high resolution optical spectrum of NGC~6302 provided by the ESO Science Archive Facility. In this paper, we investigate
the physical properties of \hi in NGC~6302 using these features.
The paper is organized as follows. In Section~2, we briefly explain the basic atomic physics of Raman scattering with atomic hydrogen. 
In Section~3, we analyze Raman-scattered \heii features  at 4851 \AA\ as well as at 6545 \AA.
We present the results from our Monte Carlo simulations in Section~4. A brief summary
and discussions are presented in the final section.

\begin{figure*}[ht!]
\plotone{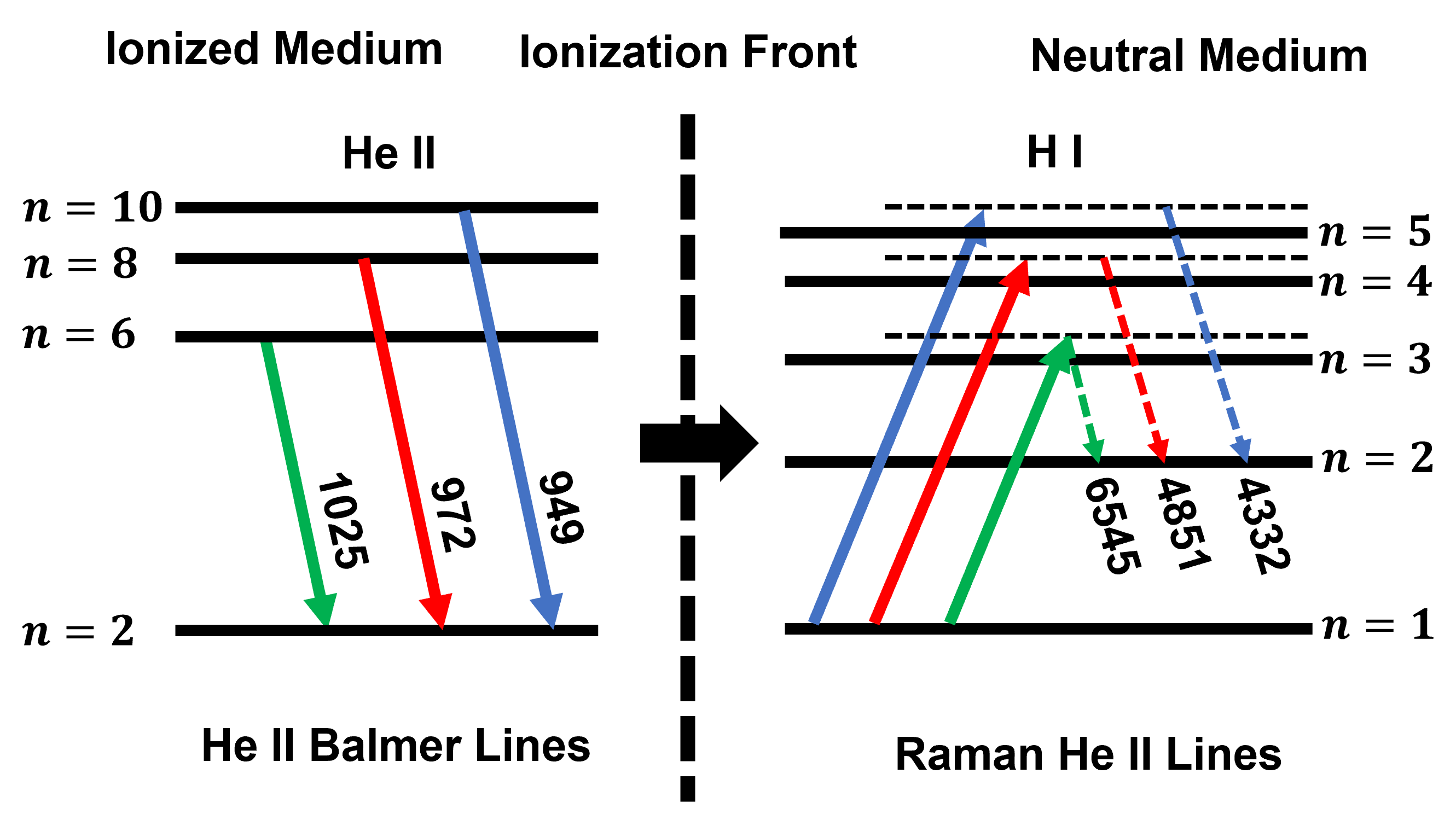}
\caption{A schematic illustration to show the formation of Raman-scattered \heii features
blueward of \hi Balmer lines in the \hi region neighboring the \heii emission region.
Far UV \heii lines at 1025, 972 and 949 \AA\ are slightly more energetic than \hi Ly$\beta$,
Ly$\gamma$, and Ly$\delta$, resulting in optical lines at 6545 \AA, 4851 \AA, and 4332 \AA\ 
blueward of H$\alpha$, H$\beta$ and H$\gamma$.
\label{fig:raman_schematic}}
\end{figure*}

\section{Atomic Physics}

\cite{schmid89} proposed that the broad emission features at 6825 \AA\ and 7082 \AA\
that appear in about a half of symbiotic stars are formed through inelastic scattering
processes of the far UV resonance doublet lines \ion{O}{6}$\lambda\lambda$1032 and 1038.
When a far UV photon blueward of Ly$\alpha$
incident on a hydrogen atom in the ground state can be converted into a lower energy photon
by an amount of 10.2 eV, the energy of a Ly$\alpha$ photon, then the hydrogen
atom makes a final de-excitation into the excited $2s$ state instead of the initial ground state.

Additional examples are provided from far UV \heii lines. \heii$\lambda$1025 arising from transitions of $6\to2$ is 
Raman scattered with atomic hydrogen to form an optical
spectral feature at 6545 \AA, blueward of \heii$\lambda$6560, which is associated
with transitions $6\to4$. Similarly, Raman scattering of  \heii$\lambda$972 and $\lambda949$ yields
spectral features at 4851 \AA\ and 4332 \AA, respectively \citep{nussbaumer89, lee12}.

Figure~\ref{fig:raman_schematic} shows a schematic illustration of the Raman scattering process
expected to operate in young planetary nebulae, 
when far UV \heii line photons enter the neutral region behind the ionization front.
The atomic line center wavelengths for far UV \heii emission lines and their optical Raman lines
are shown in Table~\ref{tab:atomic}, where 
the cross sections $\sigma_{1s}^{\rm Ray}$
and $\sigma_{2s}^{\rm Ram}$ for Rayleigh and Raman scattering are also available
\citep[e.g.,][]{lee12, chang15}.

The energy conservation requires
\begin{equation}
\nu_i = \nu_o+\nu_{\rm Ly\alpha},
\end{equation}
where $\nu_i, \ \nu_o$ and $\nu_{\rm Ly\alpha}$ are frequencies of the incident, Raman-scattered
and Ly$\alpha$ photons, respectively.
One may immediately note that the line widths of the incident and the Raman-scattered radiation are related
by
\begin{equation}\label{eq:broadening}
{\Delta \nu_i\over \nu_i} = \left( {\nu_o\over \nu_i} \right) {\Delta \nu_o\over\nu_o},
\end{equation}
which shows that the Raman \heii features blueward of Balmer lines are broadened by the
factor $\nu_i/\nu_o$. 
The line broadening effect associated with the inelasticity of Raman scattering allows one to
readily identify Raman-scattered features  \citep{schmid89}.

\begin{table*}[ht!]
\centering
\caption{Atomic line center wavelength $\lambda_0$ of \heii emission and Raman \heii (first and second columns) and cross sections for Rayleigh scattering and Raman scattering into $2s$ (third and fourth columns)}
 \begin{tabular}{ccccc}
\hline
Transition     &  $\lambda_0$ of \heii Emission$^a$ [\AA]  & $\lambda_0$ of Raman \heii$^b$ [\AA ] & $\sigma^{\rm Ram}_{2s}$ [$\rm cm^{2}$]  &  $\sigma^{\rm Ray}_{1s}$  [$\rm cm^{2}$] \cr    
\hline
$n=6\to2$ & 1025.28 & 6544.70 & $ 1.2\times10^{-21} $ &  $ 6.2\times10^{-21} $ \cr
$n=8\to2$ & 972.13 &4851.30 & $1.4\times10^{-22} $ &  $8.3\times10^{-22} $ \cr
$n=10\to2$ & 949.32 &4331.74 & $2.9\times10^{-23} $ &  $1.9\times10^{-22} $ \cr
\hline
$^a$ vacuum wavelength, $^b$ air wavelength
 \end{tabular}
 \label{tab:atomic}
\end{table*}

\section{Observation}

\subsection{Data}

\begin{figure*}[ht!]
\epsscale{1.15}
\plotone{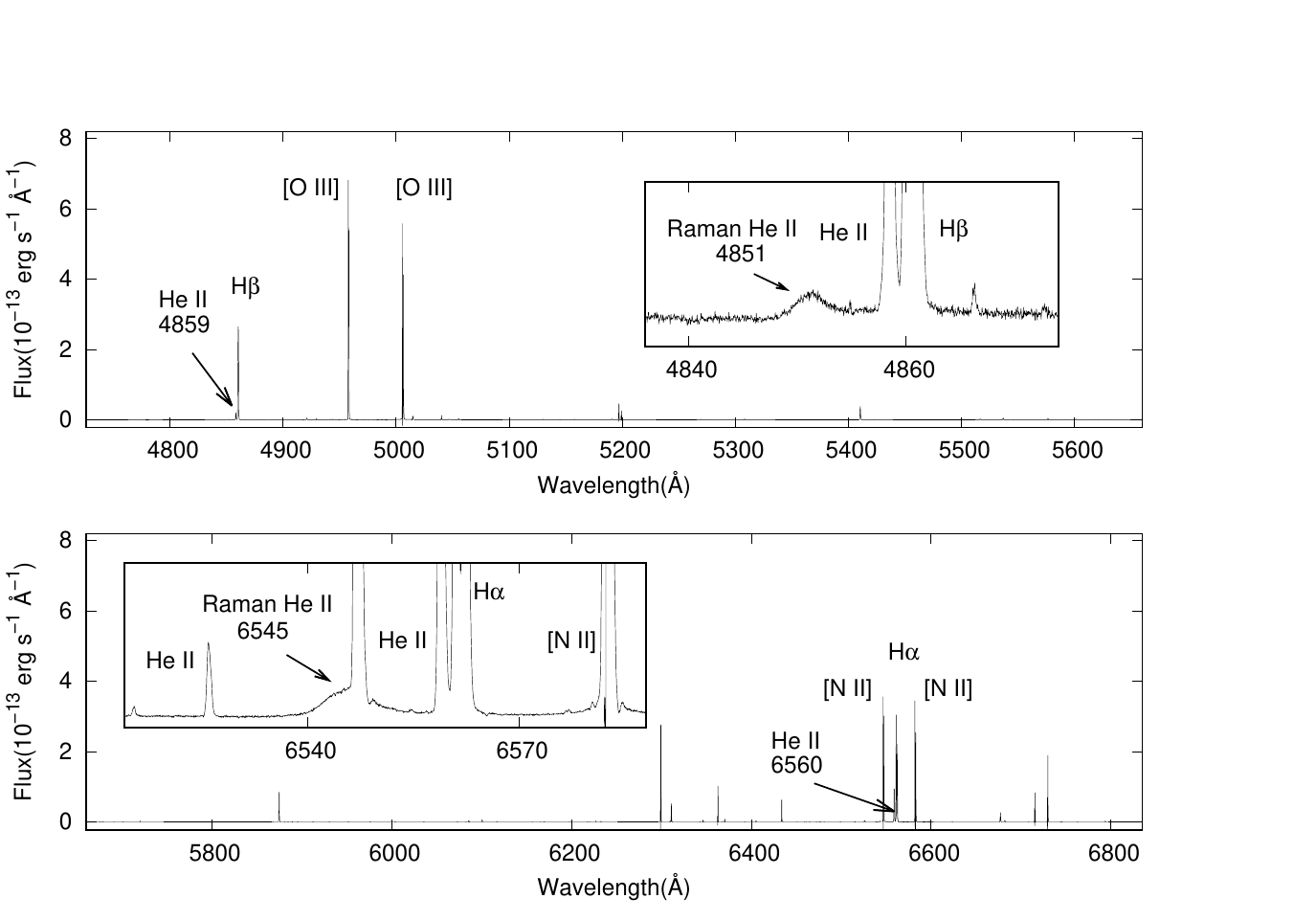}
\caption{The optical spectrum of NGC~6302 retrieved from the UVES ESO public data archive. 
Strong emission lines, including H$\alpha$, H$\beta$, [O~III]$\lambda\lambda$4959, 5007, and [N~II]$\lambda\lambda$6548, 6583 are heavily saturated and \heii emission lines at 4859 \AA\ and 6560 \AA\ are barely 
noticeable. Due to the heavy saturation, [\ion{N}{2}]$\lambda$6548 appears comparable to
[\ion{N}{2}]$\lambda$6583, even though [\ion{N}{2}]$\lambda$6548 is in fact 3 times weaker than
[\ion{N}{2}]$\lambda$6583. The insets of the upper and lower panels zoom in spectral regions around H$\beta$
and H$\alpha$, respectively. In the insets, the two broad features at 4851 \AA\ and 6545 \AA\ are clearly seen, which are Raman-scattered \heii features.
\label{fig:full_spect}}
\end{figure*}

We retrieved the spectra of NGC~6302 taken with UV-Visual Echelle Spectrograph (UVES) attached on ESO Very Large Telescope (VLT) from ESO Science Archive\footnote{http://archive.eso.org/scienceportal/home} under the ESO programme 65.I-0465 (P.I. Casassus). 
NGC~6302 was observed on May 24, 2000 through a 0.3 arcsec slit, resulting in spectral resolution of $R\sim 107,200$. 
A total of 4 spectra were obtained. The exposure times are 1200 seconds for the first 2 spectra and 2400 seconds for the other 2 spectra.
The median signal-to-noise ratio of each exposure ranges from 7.1 to 11.5.

The 1-D and wavelength-calibrated spectrum covers the spectral range from 4727 \AA\ to 6835 \AA. In Figure~\ref{fig:full_spect}, we show the full UVES spectrum of NGC~6302. The strong lines of H$\alpha$, H$\beta$, [O~III]$\lambda\lambda$4959, 5007
and [N~II]$\lambda\lambda$6548, 6583 are all severely saturated in this spectrum.
The presence of \heii emission lines at 6560 \AA\ and 4859 \AA\ is barely noticed in the scale.

In the inset of the upper panel of Figure~\ref{fig:full_spect}, we show a part around H$\beta$ 
 to show clearly the broad feature at 4851 \AA, which is identified as Raman-scattered \heii. In a similar way, the inset of the lower panel
shows a part of the spectrum near H$\alpha$, where we
find a broad feature blended with [N~II]$\lambda$6548. 
The intrinsic flux ratios
$F_{6583}/F_{6548}$ of [\ion{N}{2}]$\lambda\lambda6548, 6583$ as well as
$F_{5007}/F_{4959}$ of [\ion{O}{3}]$\lambda\lambda4959, 5007$
is fixed to be 3 due to the fact that they arise from the same excited state. Hence, if the broad weak
emission feature near [\ion{N}{2}]$\lambda$6548 is due to \ion{N}{2}, then it implies the presence of a more conspicuous broad feature by a factor 3 near [\ion{N}{2}]$\lambda$6583.
No such broad feature is apparent
near [N~II]$\lambda$6583, 
which confirms that the broad feature around [N~II]$\lambda$6548
is not associated with [N~II] but is to be identified with Raman-scattered \heii.

\subsection{Gaussian Fitting}

In this subsection, we provide a quantitative profile analysis of \heii emission
lines and Raman-scattered \heii lines. In addition to \heii$\lambda$6560 and \heii$\lambda$4859, we also consider \heii$\lambda$6527, which is formed from transitions between $n=14$ and $n=5$. This emission line is very convenient
because Raman \heii $\lambda$6545 is located redward of \heii~$\lambda$6527 with comparable strength.
The atomic line centers of these \heii $\lambda$4859, $\lambda$6527 and $\lambda$6560 are 4859.32 \AA, 6527.10 \AA, 6560.10 \AA, respectively \citep[e.g.,][]{lee06,hyung04}.

The \heii emission lines and Raman-scattered features of NGC~6302 appear to be well-fitted using a single 
Gaussian function in the form given by
\begin{equation}
    G(\lambda) = {F \over \sqrt{2 \pi}\ \sigma_{\lambda}} \exp \left(-{ (\lambda-\lambda_c)^2 \over {2 \sigma_{\lambda}^2} } \right),
\end{equation}
where $\lambda_c$ and $\sigma_{\lambda}$ are the wavelength of line center and the line width, respectively.

\begin{figure*}[ht!]
\epsscale{1.15}
\plotone{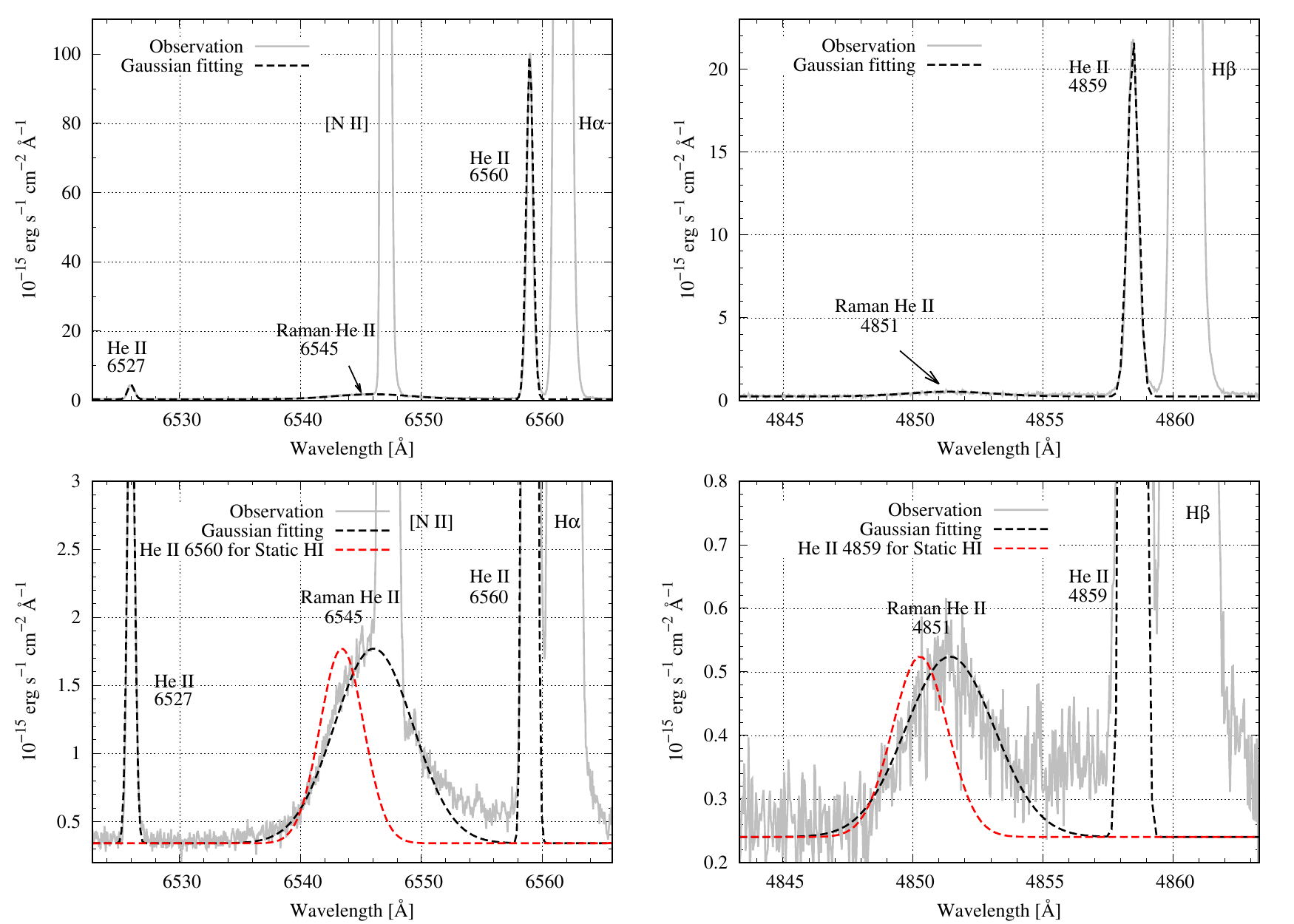}
\caption{ Gaussian line profile fits of NGC~6302 near H$\alpha$ (left panels) and H$\beta$ (right panels). 
The observed data are represented by the gray solid lines. 
The black dashed lines show the results of our single Gaussian fit adopting parameters in Table~\ref{tab:fitting}. 
In the lower two panels, the red dashed lines show
hypothetical Raman-scattered features that would be formed in a \hi region stationary with
respect to the \heii emission region.
The observed Raman \heii are broader and more red-sfhited than the hypothetical profiles. 
\label{fig:observation}}
\end{figure*}

In Figure~\ref{fig:observation}, we show the result of our single Gaussian fits. The observational data are shown with the solid gray lines, and the Gaussian fitting results are 
shown with the black dashed lines. In the two left panels, the results for \heii $\lambda$6527 and $\lambda$6560 and Raman \heii $\lambda$6545 are shown, and the right panels show the results for the counterparts blueward of H$\beta$. 
The left lower panel shows the Raman \heii $\lambda$6545, which is severely blended with [N II]$\lambda$6548. The right lower
panel shows the Raman \heii $\lambda$4851.

Table~\ref{tab:fitting} provides the fitting parameters used for the results that are shown with the black dashed lines in Figure~\ref{fig:observation}.
The widths $\sigma_\lambda$ of the two \heii emission lines $\lambda$6560 and $\lambda$6527 near H$\alpha$ are 0.283 \AA,
and that of \heii$\lambda$4859 is 0.212 \AA. In contrast, the line width
$\sigma_\lambda=3.25 {\rm\ \AA\ }$ of Raman \heii $\lambda$6545 is 11.5 times wider than
those of \heii $\lambda$6527 and $\lambda$6560. In the case of Raman
\heii $\lambda$4851, the line width $\sigma_\lambda=1.70$ \AA\ is wider than that of \heii$\lambda$4859
by a factor of 8.02. These factors exceed the line broadening factor $\lambda_o/\lambda_i$ due
to the inelasticity of Raman scattering given in Eq.~(\ref{eq:broadening}).
This is consistent with the proposal that the Raman-scattered \heii features of NGC~6302 are formed in
expanding neutral regions, where the expansion provides additional line broadening \citep[e.g.][]{jung04, choi20a}.

\begin{table*}[ht!]
\centering
\caption{Parameters of Gaussian fitting in Figure~\ref{fig:observation}}
 \begin{tabular}{cccccc}
\hline
Line                    &  \heii 6560  & Raman \heii 6545 &  \heii 6527   & \heii 4859   & Raman \heii 4851  \cr    
\hline
$\lambda_c$ [\AA] & 6559.00         &   6546.06     & 6525.98   &  4858.45 & 4851.45    \cr
$\sigma_{\lambda}$ [\AA] & 0.283         &   3.25     & 0.283   &  0.212 & 1.70    \cr
F [$10^{-14}\rm erg \, s^{-1} \, cm^{-2}$] & 7.08 & 1.17 & 0.292 & 1.16 & 0.121 \cr  
$\sigma_{v}$ [\kms] & 12.9 & 23.4$^{a}$ &  12.9 & 13.09 & 21.03$^{a}$ \cr
$\Delta V_{c}$ [\kms] & - & 17.86$^{b1}$ &  - & - & 12.88$^{b2}$ \cr
\hline
 \end{tabular}
 \footnotesize{a:~corrected~for~Raman~broadening, b1 \& b2: velocity offset from \heii$\lambda$6560 \& $\lambda$4859}
 \label{tab:fitting}
\end{table*}

\subsection{Line Ratio of \heii $\lambda$6560 \& $\lambda$4859}\label{sec:dust}

The observed flux ratio of H$\alpha$/H$\beta$ is regarded as an extinction
indicator, because the Case~B recombination theory predicts the flux ratio of $\sim 2.8$
\citep[e.g.][]{osterbrock89, storey95}.
However, H$\alpha$ and H$\beta$ in the UVES spectra are saturated near the line center, preventing
one from estimating the extinction.
Instead, the nearby emission lines \heii $\lambda$6560 and $\lambda$4859 can be used 
for the same purpose. 

According to Case B recombination theory for \heii, the intrinsic line ratio
of \heii $\lambda$6560 and $\lambda$4859 is $\sim 2.5$, whereas it is observed to be
$\sim 6.12$ from the UVES spectra \citep{storey95}.
%
By adopting the
dust model of our Galactic interstellar medium provided by \cite{draine03},
the observed line ratio of \heii$\lambda$6560 and $\lambda$4859 is consistent with 
$N_{\rm H} = 1.35 \times 10^{22} \unitNHI$ corresponding to $A_V \sim 7.5$.

\subsection{Line Broadening}

Raman \heii lines are broadened significantly due to inelasticity
of Raman scattering, which is illustrated in Eq.~(\ref{eq:broadening}). 
Therefore, a correction
factor of $\lambda_{\rm UV} / \lambda_{\rm Ram}$ is required in order to yield the velocity width in the parent UV
spectral space, where $\lambda_{\rm UV}$ and $\lambda_{\rm Ram}$ are the wavelengths of the incident UV \heii emission and the corresponding Raman scattered feature, respectively.
The velocity width $\sigma_v$ of Raman-scattered \heii is 
calculated by
\begin{equation}
    \sigma_v = { {c \sigma_{\lambda}} \over {\lambda_c} } {\left( \lambda_{\rm UV} \over \lambda_{\rm Ram} \right) }
    = \sigma_v^{\rm app} {\left( \lambda_{\rm UV} \over \lambda_{\rm Ram} \right) },
\end{equation}
where $\sigma_v^{\rm app}=c\sigma_\lambda/\lambda_c$ is an apparent velocity width of Raman \heii. 
For example, the apparent width $\sigma_\lambda = 3.25 {\rm\ \AA}$ for Raman \heii$\lambda$6545 is converted to $\sigma_v = 23.4{\rm\ km\ s^{-1}}$ instead
of $\sigma_v^{\rm app} = 150{\rm\ km\ s^{-1}}$.

Our single Gaussian fit analysis shows that $\sigma_v=23.4{\rm\ km\ s^{-1}}$ and $21.0{\rm\ km\ s^{-1}}$
for Raman \heii $\lambda$6545 and
 $\lambda$4851, respectively.
These values of $\sigma_v$ for Raman \heii lines are broader than those of \heii emission lines by $8-10\ \kms$. 
In the lower two panels of Figure~\ref{fig:observation}, we show the hypothetical profiles of Raman scattered features, which would be formed in a static \hi medium with $\sigma_v = 13 \kms$. 
In Section~\ref{sec:monte_carlo}, we use a Monte Carlo approach to find that the difference of $\sigma_v$ between Raman \heii and \heii emission lines originates from the random motion of scattering medium.

\subsection{Line Center Shift}

In the bottom row of Table~\ref{tab:fitting}, 
we show the velocity offset $\Delta V_c $ of  Raman \heii lines relative to nearby optical
\heii emission lines. 
The apparent velocity line shift $V_{\rm app}$ is given by
\begin{equation}
 V_{\rm app} = \left( {\lambda_c - \lambda_0} \over \lambda_0 \right)c
\end{equation}
where $\lambda_{0}$ is the atomic line center wavelength.
In turn, the velocity offset
$\Delta V_c$ of Raman \heii$\lambda$6545 is calculated  by
\begin{equation}
    \Delta V_{c} = {\lambda_{0,1025} \over \lambda_{0,6545}}V_{\rm app,6545} - V_{\rm app,6560}.
\end{equation}
The wavelengths $\lambda_{0,1025},\lambda_{0,6545}$, and $\lambda_{c,6545}$ are
defined analogously.
In the case of Raman \heii$\lambda$4851, the velocity offset $\Delta V_c$ is obtained with $\lambda_{0, 972}$ and $\lambda_{0,4859}$ replacing $\lambda_{0, 1025}$ and $\lambda_{0,6560}$, respectively.

From our line fit analysis, $\Delta V_c = 10.24$ \kms and 13.10 \kms for Raman \heii$\lambda$6545 and $\lambda$4851, respectively. 
In the bottom panels of Figure~\ref{fig:observation}, 
the Raman \heii features are clearly displayed redward of the hypothetical features shown in the red dashed lines.
This redward line shift of Raman \heii indicates that the \hi medium is moving away 
from the \heii emission region
 \citep[e.g.][]{choi20a}.

\subsection{Raman Conversion Efficiency}\label{sec:RCE}

In this subsection, we present the Raman conversion efficiency (RCE), which is defined as the ratio of the number of Raman-scattered photons divided by that of the incident far UV photons.
Explicitly, RCE of Raman \heii $\lambda$6545 is given by
\begin{eqnarray}
 \nonumber
    {\rm RCE}_{6545} &=& {{F_{6545} / E_{6545}} \over {F_{1025} /E_{1025}} } 
    \\
    &=&  \left( 
\lambda_{0, 6545} \over \lambda_{0, 1025} \right) {\left({F_{6545}} \over {F_{6560} } \right)} \left( F_{6560} \over F_{1025} 
\right),
\label{eq:RCE}
\end{eqnarray}
where $E_\lambda=hc/\lambda$ is the energy of a photon with wavelength $\lambda$.
RCE of Raman \heii $\lambda$4851 is also given in an analogous way by replacing 1025 and 6560
by 972 and 4859.

Firstly, the ratio of Raman and neighboring optical \heii line fluxes, $({F_{6545}}/{F_{6560} })$ or $({F_{4851}}/{F_{4859} })$, is obtained using the observed values presented in Table~\ref{tab:fitting}.
Secondly, the flux ratio of the neighboring optical \heii and directly unavailable incident far UV \heii lines, $( F_{6560} / F_{1025} )$ or $( F_{4859} / F_{972} )$, is deduced, assuming that Case B recombination  is valid. 
According to \cite{storey95},   $( F_{6560} / F_{1025} ) \sim 0.19$ and
also $( F_{4859} / F_{972} ) \sim 0.19$. Combining these results,
we obtain RCE$_{6545}=0.21$ and RCE$_{4851}=0.10$.

\begin{figure*}
\plotone{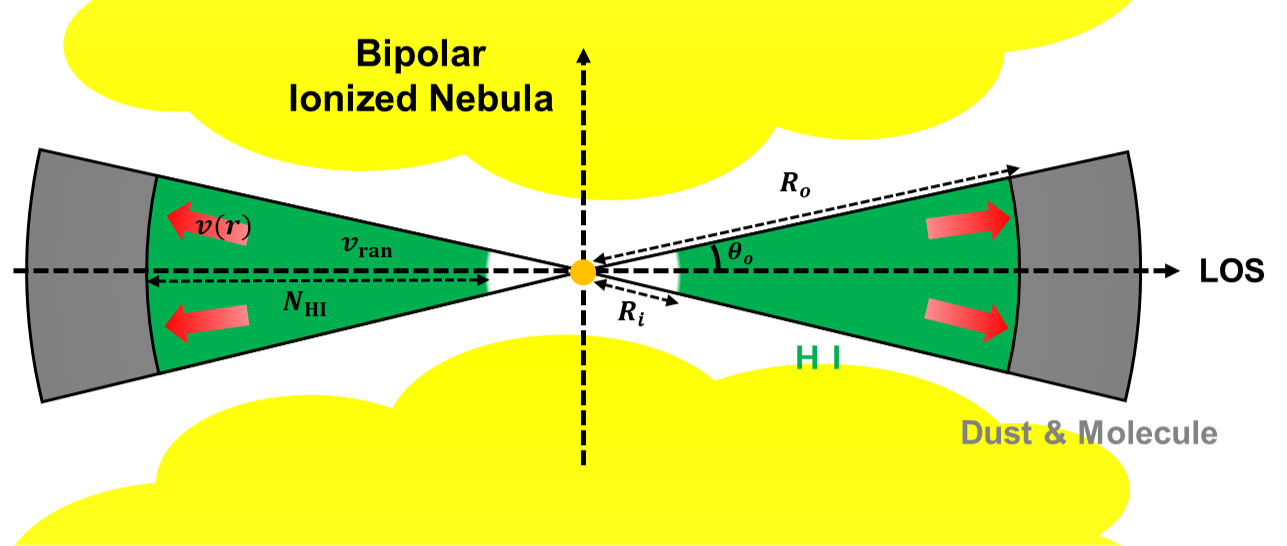}
\caption{Schematic illustration of the scattering geometry considered in this work for Monte
Carlo radiative transfer. The scattering geometry is composed of a point-like central \heii emission source (orange), a disk-like \hi region (green), and an outer dusty and molecular region (gray).
Closely related to the line shift and broadening of Raman \heii, the \hi region has two kinematic components: one is a radial expansion with a velocity $\vexp$, and the other is a random motion with a representative speed of \vran. 
The scattering geometry is specified by the inner and outer radii $R_i$, $R_o$ and the half opening angle $\theta_o$ of the \hi region with respect to the central source.
\label{fig:scat_geo}}
\end{figure*}

\section{Monte Carlo Approach}\label{sec:monte_carlo}

\begin{figure*}[ht!]
	\epsscale{1.15}
\plotone{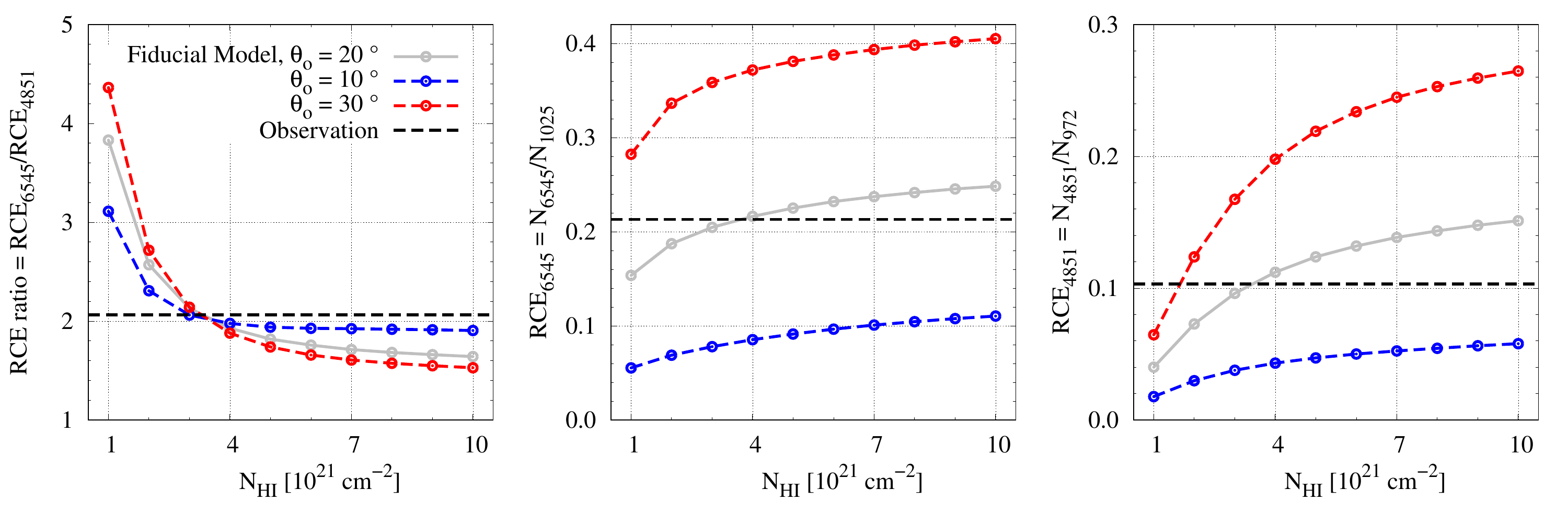}
\caption{Raman conversion efficiency (RCE) of Raman He~II 4851 (right) and 6545 (center) lines and the ratio of RCE of those two Raman He~II lines in our model of Figure~\ref{fig:scat_geo} for various half opening angle \openangle. The black vertical lines represent the values from Gaussian fitting in Table~\ref{tab:fitting}. 
\label{fig:mc_res}}
\end{figure*}

Raman \heii lines carry important physical information of the \hi region.
The line center, profile width and the strength of Raman-scattered \heii can be used to put
strong constraints on the distribution and kinematics of the \hi component near the \heii
emission nebula \citep[e.g.,][]{nussbaumer89}.  
Due to the difference in scattering cross sections of \heii $\lambda$ 1025 
and \heii$\lambda$972, the Raman conversion efficiencies also differ so that much detailed information
can be obtained if both Raman-scattered features are secured with sufficient data quality. In particular,
\cite{choi20a} investigated the line formation of Raman-scattered \heii in an expanding \hi region to show
that the redward shift of the line center of Raman-scattered \heii features is conspicuously enhanced 
due to the sharp rise of the cross section toward \hi Lyman line centers \citep[see also][]{jung04}. 
In this section, we adopt a Monte Carlo approach to propose a simple scattering geometry 
consistent with the observed spectrum considered in this work.

The Monte Carlo code '{\it STaRS}' developed for radiative transfer in thick neutral regions by \cite{chang20} is used to simulate the formation of Raman-scattered \heii and find
the best fitting parameters. In this simulation, we set the numbers of the incident far 
UV \heii 1025 and 972 photons to be $10^8$. 
The ratio of the two UV lines is $\sim 2.5$ in accordance with Case B recombination.
Because optical Raman-scattered \heii features are subject to dust extinction after
leaving the \hi region, the Raman conversion efficiency is computed using the optical Raman
photons before dust extinction. However, the final line fitting of Raman-scattered \heii is carried out after correction for
dust extinction. The incident far UV \heii photons are assumed to be described by a single
Gaussian profile with a velocity width $\sigma_v = 13{\rm\ km\ s^{-1}}$ corresponding to the widths of \heii $\lambda$6560 and $\lambda$4859 as illustrated in Table~\ref{tab:fitting}.

    \subsection{Scattering Geometry}

The central region of NGC~6302 is highly obscured and can be probed with high angular resolution
observations achievable by radio interferometry. \cite{peretto07} investigated the kinematics
of the molecular torus to report the expansion velocity of $8{\rm\ km\ s^{-1}}$. \cite{wright11}
presented their 3D photoionization computation to propose that the inner and outer radii of the circumstellar disk are $r_{\rm in}=1.2\times 10^{16}{\rm\ cm}$ and $r_{\rm out}=3.0\times10^{17}{\rm\ cm}$, respectively, based on their best-fitting model result. Furthermore, the circumstellar
disk is geometrically thin with a half-opening angle $\sim 10^\circ$.

 In our Monte Carlo simulation,
the \heii emission region is assumed to be an unresolved compact source surrounded by a circumstellar disk, where Raman scattering takes place. 
Figure~\ref{fig:scat_geo} shows a schematic illustration of the scattering geometry considered in this work.
For the sake of simplicity, the neutral
region is assumed to take the form of a disk with the inner and outer 
radii $R_{\rm i}=10^{16}{\rm\ cm}$ and $R_o=5R_{\rm i}$ and also characterized by 
the half opening angle $\theta_o$ of the \hi disk with respect to the point-like central \heii region.
The neutral region is of uniform \hi density $n_{\rm HI} = \NHI/(R_o-R_i)$, where \NHI is \hi column density. 

The \hi medium is assumed to move away from the central \heii emission
region in the radial direction. In addition,
we denote by \vran the random motion component contributed by the thermal
motion. 
A Hubble-type outflow is chosen in this work in accordance with the observations of the ionized and molecular components by \cite{szyszka11} and \cite{santander_garcia17}, respectively. 
Specifically, the radial velocity $v(r)$ at a distance $r$ from the \heii source is chosen to follow
\begin{equation}
 v(r) = \vexp \left( {r \over R_o} \right),
\end{equation}
where the parameter \vexp is the expansion velocity at the outer radius $R_o$.

In our simulation, we collect Raman scattered photons escaping along the line of sight, which coincides with the direction specified by the polar angle $\theta$= $90^\circ$, in view 
of the fact that the central star of NGC~6302 is highly obscured by dust \citep{kastner22}.
Thus, we consider dust extinction in the line of sight, which is coincident with the equatorial direction of the scattering medium.
In Figure~\ref{fig:scat_geo}, a dust component is added outside the neutral region so that optical Raman
\heii photons are subject to dust extinction before reaching the detector. The dust optical
depth is chosen to be consistent with the reddening found in the line ratio of
\heii$\lambda$6560 and $\lambda$4859 discussed in Section~\ref{sec:dust}.
The central \heii emission region is assumed to inject far UV \heii line photons with a profile described by a single Gaussian and strengths in accordance with the Case B recombination theory.

\subsection{Simulated Raman Conversion Efficiency}

\begin{figure*}[ht!]
	\epsscale{1.15}
\plotone{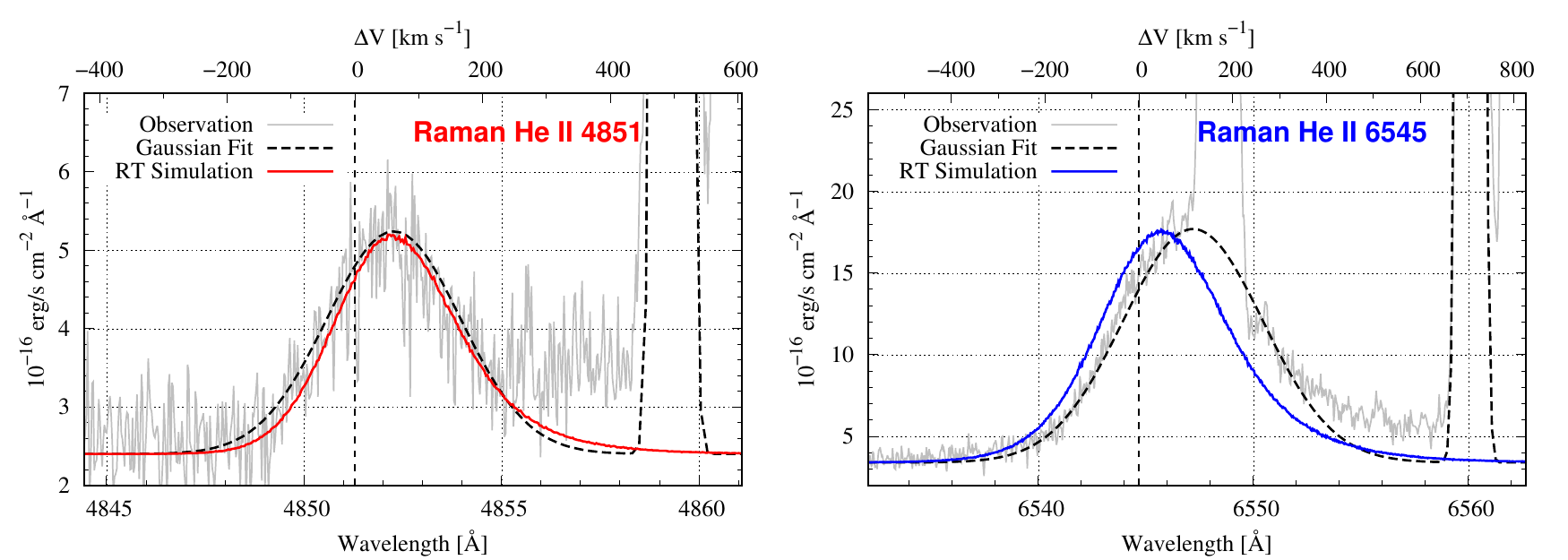}
\caption{Best fit simulation profiles of Raman \heii 4851 (left) and 6545 (right) superimposed on the observed spectrum (gray) and the single Gaussian fits (black). The simulated profiles are shown by red and
blue solid lines in the left and right panels, respectively. The model parameters are $\theta_0=20^\circ$,
$N_{\rm HI} = 3\times 10^{21}{\rm\ cm^{-2}}$, $\vexp = 13 \kms$, and $\vran = 10 \kms$.
\label{fig:mc_spec}}
\end{figure*}

In Figure~\ref{fig:mc_res}, the Raman conversion efficiencies for Raman-scattered
\heii at 6545 \AA\ and 4851\AA\ are shown in the middle and right panels for three values
of the half opening angle \openangle= 10, 20, and 30$^\circ$ and for a range of \hi column densities $10^{21}-10^{22} \unitNHI$. Here, we fix the expansion speed $v_{\rm exp}= 13 \kms$ and the random speed $\vran = 10\kms$ according to the best fitting result (see Appendix~\ref{sec:model_para}).
The horizontal dotted lines indicate RCEs of 0.21 and 0.10 for Raman-scattered \heii at 6545 and 4851, respectively, presented in Section~\ref{sec:RCE}. 
In the left panel, the ratio of the two Raman conversion efficiencies is shown.
The horizontal dotted line represents the ratio of the two RCEs $\sim 2.1$.

In the range of \hi column density \NHI
considered in Figure~\ref{fig:mc_res}, the Raman conversion efficiency is nearly
proportional to the half opening angle $\theta_0$. However, in contrast, 
the ratio of the two Raman conversion efficiencies
is relatively insensitive to $\theta_0$ and decreases as \NHI increases.
In this range of \NHI, too small RCEs are obtained from the cases for \openangle = $10^\circ$ and those with \openangle = $30^\circ$ lead to much larger RCEs to account for the measured values.  
In view of this, we propose that the measured RCEs are consistent with 
the scattering geometry characterized by
$\NHI = 3 \times 10^{21} \unitNHI$ and \openangle = $20^\circ$.

\subsection{Best Fit Profiles}

In the left panel of Figure~\ref{fig:mc_spec}, we show the best line fit to the observed spectrum of Raman \heii$\lambda$4851, for which the parameters adopted are \vexp =  13\kms and \vran = 10 \kms
in addition to $N_{\rm HI}=3\times 10^{21}{\rm\ cm^{-2}}$ and $\theta_{\rm o}=20^\circ$.
The red line shows the best fit and the black dotted line is the Gaussian fit to the observed data.

Using our best fit parameters, the \hi number density is estimated to
$n_{\rm HI}= 7.5\times 10^4{\rm\ cm^{-3}}$, from which we estimate the total mass of the \hi disk $\simeq 1.0\times 10^{-2} {\rm\  M_\odot}$. \cite{taylor90} investigated the \hi mass of several 
planetary nebulae from 21 cm radio observations. For example, they propose that the planetary 
nebula BD~+~30$^\cdot$3639 has the \hi mass of $0.028{\rm\ M_\odot}$.
The random speed of \hi medium $\sim 10 \kms$ corresponds to the thermal speed at $T = 5000 \rm\ K$, which is also consistent with the excitation temperature deduced from 21 cm radio observations.

We use the same set of parameters to show the fitting result for the Raman \heii $\lambda$6545 by the blue line in the right panel. 
The fit quality of Raman \heii $\lambda$6545 is slightly poorer than of Raman \heii at 4851 \AA. More specifically, the simulated profile is shifted blueward of the observed data by an amount
of $\sim 6{\rm\ km\ s^{-1}}$.
Because of the severe blending of Raman \heii $\lambda$6545 with \ion{N}{2}$\lambda$6548, we consider it a more appropriate strategy to
focus on the Raman \heii $\lambda$4851 in the determination of the best fit parameters
pertinent to NGC~6302.
In Appendix~\ref{sec:model_para}, we show the dependence of spectral profile on parameters, demonstrating the best fit convincingly.

\section{Summary and Discussion}

Using the archived UVES spectrum of NGC~6302, we have carried out
profile analyses of \heii emission lines and Raman-scattered \heii at 6545 \AA\ and 4851 \AA.
Raman \heii features are found to be broader and more redshifted than the hypothetical Raman features that would be formed in a static \hi medium. 
The Monte Carlo simulation code '{\it STaRS}' \citep{chang20} is used to produce satisfactory line fits to the Raman-scattered \heii features. For our Monte Carlo approach, the \heii emission region is assumed to be compact near
the hot central star surrounded by the \hi region  with \hi column density \NHI = $3\times 10^{21} \unitNHI$ and a half opening angle $\theta_o=20^\circ$. The kinematics of the \hi region is characterized by
the expansion speed $v_{\rm exp}= 13{\rm\ km\ s^{-1}}$ and a random speed of $\vran = 10{\rm\ km\ 
s^{-1}}$.
The physical properties of \hi region are imprinted in Raman \heii features via scattering.

In this work, dust particles are assumed to be distributed outside the neutral medium, presuming that those inside the neutral region are completely destroyed by strong UV radiation from the central star.
However, additional complications are expected if we introduce a considerable amount
of dust in the \hi medium.
In a sophisticated model involving a dusty neutral medium, 
the formation of
Raman \heii $\lambda$4851 would be more suppressed than that of Raman \heii $\lambda$6545,
because \heii$\lambda$972 has the Raman cross section smaller almost by an order of
magnitude than \heii $\lambda$1025. A line photon of \heii $\lambda$972 has to traverse a longer dusty path in order to yield a Raman optical photon than \heii $\lambda$1025 so that dust extinction is more
effective for \heii $\lambda$972 than \heii $\lambda$1025.
We defer the formation of Raman features in a dusty neutral medium for future work.

On the observational side, we expect that Raman-scattered
\heii feature at 4332~\AA\ can be obtained from very deep spectroscopic observations.
Raman \heii $\lambda$4332 was reported
in the symbiotic stars, including V1016~Cygni and RR~Telescopii, and also in young planetary nebulae
including NGC~7027 \citep[e.g.,][]{lee12, vangroningen93, pequignot97}. 
With the future availability of Raman \heii $\lambda$4332, strong constraints
on the scattering geometry and the amount of dust extinction will be placed.

\begin{acknowledgments}

We are grateful to an anonymous referee for constructive comments.
This research has made use of the services of the ESO Science Archive Facility.
H.L. and J.K. were supported by the National Research Foundation of Korea (NRF) grants funded by the Korea government (NRF-2023R1A2C1006984).
\end{acknowledgments}

%

\vspace{5mm}
\facilities{ESO Science Archive Facility}





\appendix
\restartappendixnumbering

\section{Monte Carlo Profile Fitting}\label{sec:model_para}

\begin{figure*}[ht!]
	\epsscale{1.15}
\plotone{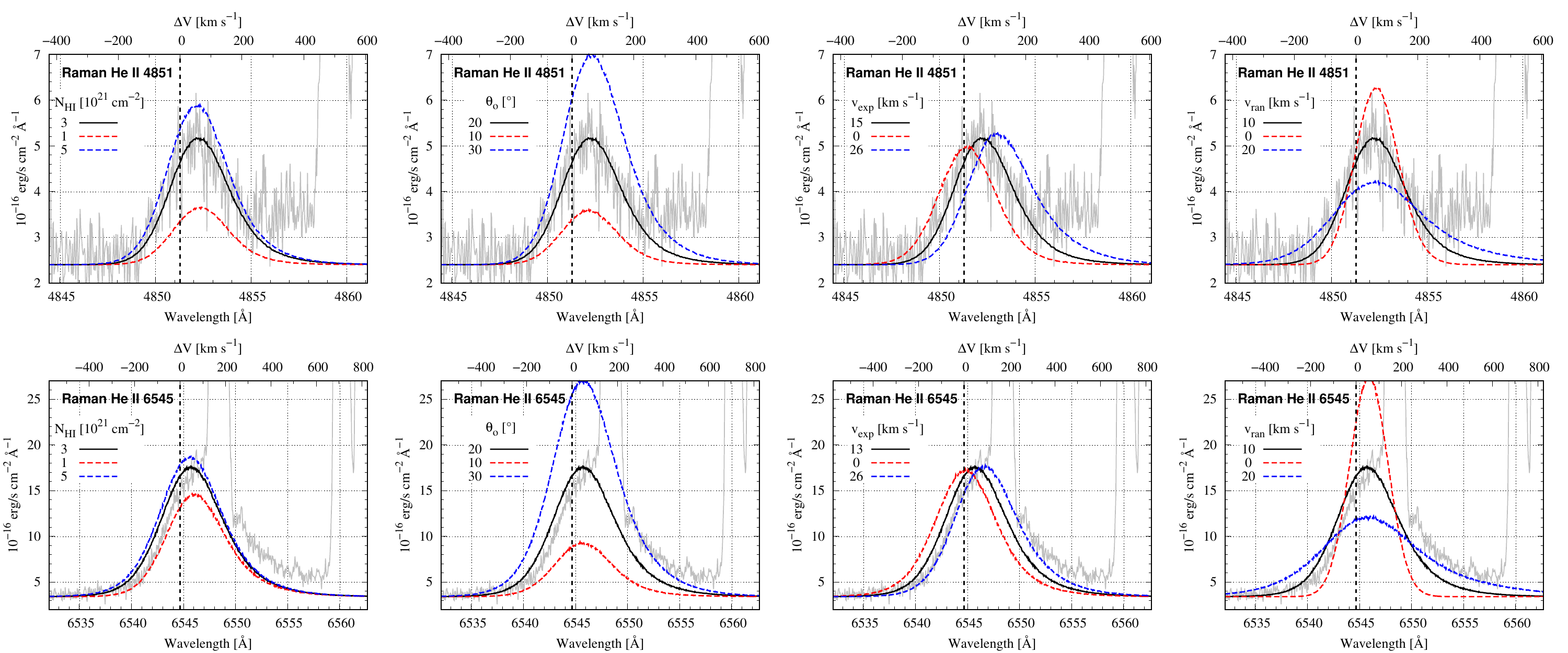}
\caption{ Fitting results using Raman \heii 4851 (top) \& 6545 (bottom) spectra from {\it STaRS}. The gray and black solid lines are observed spectra and the best-fitting results, respectively. The red-blue solid dashed lines represent the spectra for smaller-larger parameters (\NHI, $\theta_{\rm o}$, \vexp, and \vran).}
\label{fig:mc_par}
\end{figure*}

\begin{figure*}[ht!]
	\epsscale{1.15}
\plotone{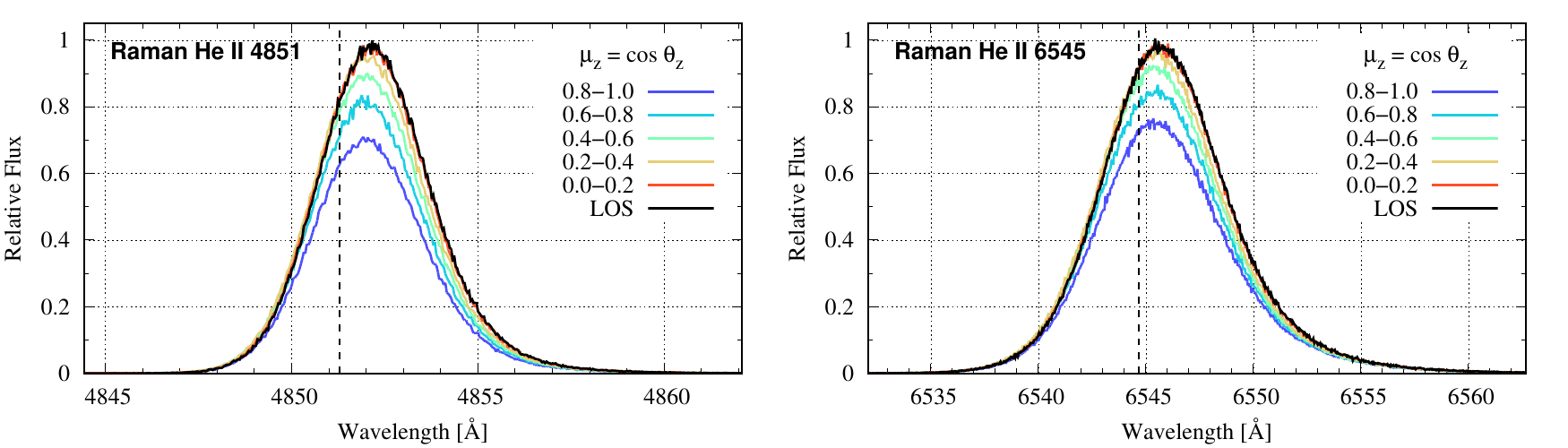}
\caption{ Simulated spectra of Raman \heii $\lambda$4851 (left) and $\lambda$6545 (right) for various escaping directions without dust extinction.
The colors represent the different directions $\mu_z = \cos\theta_z$, where $\theta_z$ is the polar angle measured from the symmetry axis, $z$-axis. Raman-scattered features are observed to be weaker for $\mu_z = 0.8-1.0$ than for $\mu_z=0.0-0.2$ mainly due to the anisotropic scattering phase function favoring the
forward and backward scattering.
The black lines are identical to the best-fit spectra shown in Figure~\ref{fig:mc_par}.}
\label{fig:mc_los}
\end{figure*}

The expanding velocity and random speed of \hi are mainly responsible for the center shift and line broadening of Raman \heii, respectively. However, the line profiles of Raman-scattered \heii
are determined in a complicated way involving the multiple scattering effect and sharply varying cross
sections as a function of wavelength in addition to the scattering geometry and the kinematics.
For example, when the covering factor of the neutral scattering region is significant, far UV \heii 
photons that escape through Rayleigh scattering may re-enter the scattering region, introducing
an additional line broadening effect to the final Raman-scattered \heii \citep[e.g.,][]{choi20a}.
However, the half opening angle of $20^\circ$ deduced from Figure~\ref{fig:mc_res} is not big enough to enhance
the line broadening through the re-entry effect.
For this reason, we consider the random speed component of the \hi medium in our simulation as the main
factor affecting the broadened Raman-scattered \heii.

In Figure~\ref{fig:mc_par}, we show line profiles obtained from our Monte Carlo simulations adopting the scattering
geometry illustrated in Figure~\ref{fig:scat_geo}. In particular, we show the dependence
of the Raman line profiles on the parameters a \hi column density \NHI, an half opening angle \openangle, an expansion velocity \vexp, and a random speed \vran. The upper and lower panels show the results for Raman \heii $\lambda$4851 and  Raman \heii $\lambda$6545, respectively.

In the two left panels, we investigate
the effect of \NHI, where the three cases for $\NHI/(10^{21}{\rm\ cm^{-2}})
=1,3$ and 5 are shown. The other parameters are fixed to the best fit values, i.e., $\openangle=20^\circ$, $\vexp = 13 \kms$ and $\vran = 10 \kms$. Because of the best fit value $\NHI = 3\times 10^{21}{\rm\ cm^{-2}}$
corresponding to Raman optical depth exceeding unity for \heii$\lambda$1025,
the Raman conversion efficiency increases only slightly compared to that for \heii$\lambda$972.

The next two panels show the effect of \openangle, where the three
values of $\openangle = 10^\circ,20^\circ$ and $30^\circ$ are considered. The Raman conversion efficiency
is nearly proportional to \openangle, which determines the covering factor of the
scattering region with respect to the \heii emission region. 

The third two panels show the dependence of \vexp,  which mainly affects
the line center location of the Raman-scattered \heii. The line centers move redward 
as \vexp increases. It is particularly notable
that Raman \heii $\lambda$4851 becomes stronger as \vexp increases. This is
due to the increase of Raman scattering cross section as \heii photons get redshifted
toward the hydrogen resonance, while the Raman optical depth is less than unity \citep[e.g.,][]{jung04, choi20a}. No such conspicuous increase is seen in the case of Raman \heii $\lambda$6545 because the
Raman optical depth exceeds unity.

In the right two panels, the simulated line profiles for three values of $\vran = 0, 10$ and $20 {\rm\ km\ s^{-1}}$ are illustrated. 
The line profile becomes broader with increasing \vran.
Because the \heii emission region is assumed to be characterized by a random velocity $\sigma_v = 13{\rm\ km\ s^{-1}}$ corresponding to the width of \heii emission lines in Table~\ref{tab:fitting}, the resultant profiles are significantly affected by choice of \vran
in the range 0-20 \kms. In view of the profile fit quality, we may safely conclude that
the random velocity component in the neutral region is $\vran \simeq 10 \kms$.

We conclude that the best fitting parameters are obtained from the fitting of Raman \heii $\lambda$4851 and the blueward of Raman \heii $\lambda$6545.
In the third two panels,
the simulated spectra of Raman \heii $\lambda$6545 for \vexp = 13 and 26 \kms provide well-fitted features in the blueward and redward, respectively.
We set \vexp = 13 \kms  since the strong [\ion{N}{2}] emission above Raman \heii $\lambda$6545 can affect the spectral profile in the redward of 6545 \AA.

Figure~\ref{fig:mc_los} shows the simulated line spectra for various values of escaping direction $\mu_z=\cos\theta_z$, where $\theta_z$ is the angle of the emergent photon making with the symmetry
axis. The spectra for the line of sight ($\mu_z = 0$) in our simulation are stronger than those for other directions with $\mu_z\ge 0.8$. This behavior is explained by the fact that both Raman  and Rayleigh processes prefer scattering in the forward and backward directions to lateral directions, where
the phase function of the two scattering processes is given by
\begin{equation}
    \Phi(\mu_s) = {3 \over 8}( 1 + \mu_s^2).
\end{equation}
Here, $\mu_s = {\bf\hat k_i}\cdot {\bf\hat k_s}$ is the cosine of the angle between 
the wavevectors ${\hat k_i}$ and ${\hat k_s}$  of the incident and scattered photons, respectively
\citep{chang20}. 





\bibliography{references}{}
\bibliographystyle{aasjournal}



\end{document}